\documentclass[11pt]{article}

\def\beq{\begin{eqnarray}}
\def\eeq{\end{eqnarray}}

\def\al{\alpha}
\def\be{\beta}

\def\ga{\gamma}\def\de{\delta}

\def\ep{\epsilon}

\def\la{\lambda}
\def\na{\nabla}

\def\si{\sigma}
\def\om{\omega}

\def\ta{\tau}

\def\Ga{\Gamma}

\begin{document}

\title{Non-metricities, torsion and fermions}
\author{Guilherme de Berredo-Peixoto\footnote{Email address:
gbpeixoto@hotmail.com} \\
Departamento de F\'{\i}sica, ICE, Universidade Federal de Juiz de Fora \\ Juiz de Fora, MG, Brazil  36036-330.}
\date{April 2018}
\maketitle
\begin{abstract}
I investigate the general extension of Einstein's gravity by considering the third rank non-metricity tensor and the torsion tensor. The minimal coupling to Dirac fields faces an ambiguity coming from a severe arbitrariness of the Fock-Ivanenko coefficients. This arbitrariness is fed in part by the covariant derivative of Dirac matrices, which is not completely determined as well. It is remarkable that this feature is not exclusive to the non-metricity case: it happens also for gravity with torsion alone. Nevertheless, theory in vacuum is well defined and non-trivial, where torsion is the source of non-metricity or vice-versa. I point also to the existence of two independent non-metricities.
\end{abstract}

PACS 04.20.-q; 04.50.Kd.

\section{Introduction}

Besides the usual Riemannian geometric degrees of freedom, non-Riemannian counterparts such as torsion \cite{hehl} and non-metricity \cite{adler,poberii} have been extensively investigated in literature as gravitation components which would be relevant in high-energy regimes. Torsion and non-metricity are respectively related mathematically to the antisymmetric part of the affine connection and the non-vanishing covariant derivative of metric tensor, and the study of their physical relevance is going on until the present days. Torsion has been considered in a greater number of works (see, for example, the reviews \cite{shapiro,saridakis}), but the interest on non-metricity have been increased in last years.

Like torsion, non-metricity can be studied in different ways according to which component of non-metricity is chosen to be considered. For example, one can consider only a scalar degree of freedom of non-metricity.\footnote{There is no scalar irreducible component in the decomposition of non-metricity tensor, such that I mean scalar degree of freedom as the unique degree of freedom coming from the vector component.} In such theories, the spacetime is called \lq\lq Weyl integrable spacetime" (see, for example, Refs. \cite{salim1,salim2,aguila}). One can consider instead the vector degrees of freedom, which brings us the Weyl spacetime, where the covariant derivative of metric tensor is proportional to the metric,
$\nabla_\mu g_{\al\be} = \phi_\mu g_{\al\be}$, and the vector $\phi_\mu$ describes all the non-metricity. If this quantity can be written as a scalar derivative, $\phi_\mu = \partial_\mu \Omega$, then the spacetime is Weyl integrable (previous case). For this approach, see, for instance, Refs. \cite{adler,fariborz,baburova,koivisto,moffat}. In the most general case, the gravity theories are called sometimes by Metric-Affine Gravity (MAG), and the spacetime is more general than the Weyl spacetime. I can cite some works in this approach in Refs. \cite{adak,formiga,olmo,koste,snow,latorre}. In each one of these three different approaches, you can find many papers (from which I cited just a few) investigating several topics, such as Dirac fermions in curved space and consequent experimental bounds, as well as cosmological effects of non-metricity like singularity avoidance or accelerated expansion. In particular, paper in Ref. \cite{koste} obtains experimental bounds for non-metricity from results already obtained for Lorentz violating theories. One finds also an increasing number of works exploring non-metricity as an equivalent to General Relativity, called Symmetric Teleparallel Gravity \cite{stg} (see also the review \cite{review}).

In the present work, I consider the most general extension of Einstein-Hilbert action, which includes torsion and general non-metricity. Firstly, in order to find how fermions couple minimally to non-metricity, one has to consider the issue of covariant derivative with respect to both diffeomorphism and Lorentz transformation, the last one understood as operating in the tangent space of each point in the manifold.\footnote{The necessity of a precise definition of how the covariant derivative with Greek
index acts upon objects with Latin indices is attenuated by observing that one can adopt $\tilde{\na}_\mu = e^a\mbox{}_\mu \tilde{\na}_ a$. Notice that the operator $\tilde{\na}_\mu$ is actually defined by acting on both types of indices. Indeed, even when one considers only Greek indices, no one can ignore that, for example, $\tilde{\na}_\mu g_{\al\be} = \tilde{\na}_\mu (e^a\mbox{}_\al e_{a\be})$, such that the complete definition of $\tilde{\na}_\mu$ is always hidden in the equations without Latin indices.} Because of the independence between spacetime and the tangent space \cite{adler}, it seems unnatural the identification of Minkowskian non-metricity,
$\tilde{\na}_\mu \eta_{ab}$, to the usual non-metricity $\tilde{\na}_\mu g_{\al\be}$. In fact, this identification lead not to any contradiction, that is why nobody has spoken about two different non-metricities. Some authors study the usual non-metricity, and another authors study the Minkowskian one. I argue that it is more natural to accept both non-metricities, mutually independent. This issue is extremely relevant for investigating how non-metricity couples to matter.

But to study the coupling to matter, one has also to understand the structure of the spinorial covariant derivative, i.e., the Fock-Ivanenko coefficients. Then, in Section 2, after setting up notations and basic features of theory, I consider gravity with torsion and the usual non-metricity in vacuum. It turns out that, in vacuum, non-metricity is the source of torsion (or vice-versa), confirming the result obtained by Ponomariov and Obukhov \cite{pono}. Thus, in principle, the detection of non-Riemannian structure in this context can face the difficulty of saying which one exists: torsion or non-metricity. In order to avoid this problem of arbitrary torsion (or non-metricity), one has to consider another action \cite{pono}.

In Section 3, the covariant derivative of Dirac matrices are calculated and restricted maximally although it remains some arbitrariness. In the next section, I express the Fock-Ivanenko coefficients in terms of the covariant derivative of Dirac matrices for the most simple choice dropping the arbitrariness but not all of them, and the standard form of the Fock-Ivanenko coefficients (present in other papers) are reproduced. In Section 4, I calculate the Fock-Ivanenko coefficients with all explicit arbitrariness, and draw my conclusions in Section 5 by observing that the same arbitrariness is also present in theory with only torsion.

\section{Formulation of theory}

We use two kinds of labels in specifying the components of a tensor: a Latin letter
($a$, $b$, ...) and a Greek letter ($\mu$, $\nu$, ...). Latin letters indicate
components of an object which is invariant under Poincar\'e transformations; and Greek
letters indicate components of an object which is invariant under general coordinate
transformations.

Let us denote the known objects (like connection and covariant derivative) with
tilde in the presence of torsion and non-metricity, and without tilde for the case
where there is only torsion. Thus, the covariant derivative of metric is written as
\beq
\tilde{\na}_\mu g_{\al\be} = \partial_\mu g_{\al\be} - \tilde{\Ga}^\rho\mbox{}_{\al\mu}g_{\rho\be} -
\tilde{\Ga}^\rho\mbox{}_{\be\mu}g_{\al\rho} = Q_{\mu\al\be}\,. \label{defQ1}
\eeq
Of course, one has, at the same time, $\na_\mu g_{\al\be} = 0$. In fact,
equation (\ref{defQ1}) is the definition of non-metricity tensor
$Q_{\mu\al\be}$ (notice that $Q_{\mu\al\be} = Q_{\mu\be\al}$). \\

Similarly, the covariant derivative of the Minkowski metric tensor has the
form\footnote{The covariant derivative of a spinor in the presence of non-metricity is a complicated subject, admitting different approaches. See, for instance, Refs. \cite{neeman,hurley,fatibene}.}
\beq
\tilde{\na}_\mu \eta_{ab} = \partial_\mu \eta_{ab} + \tilde{\om}^c\mbox{}_{a\mu}\eta_{cb} +
\tilde{\om}^c\mbox{}_{b\mu}\eta_{ac} = Q_{\mu ab}\,, \label{defQ2}
\eeq
where, obviously, $\partial_\mu \eta_{ab} = 0$ (we put it in the equation just for the sake of
completeness). The spin connection here, $\tilde{\om}_{ab\mu}$, differs from the usual spin
connection with torsion and metricity, $\om_{ab\mu}$, in respect to the known antisymmetry in
the last case, $\om_{ab\mu} = -\om_{ba\mu}$. By equation (\ref{defQ2}), the non-metricity object
$Q_{\mu ab}$ defines precisely the symmetric part of $\tilde{\om}_{ab\mu}$:
\beq
\tilde{\om}_{(ab)\mu} = \frac{1}{2}( \tilde{\om}_{ab\mu} + \tilde{\om}_{ba\mu} ) = Q_{\mu ab} \,.
\label{symspinconnection}
\eeq

So far we have defined two kinds of non-metricity, $Q_{\mu\al\be}$ and $Q_{\mu ab}$, without indicating
any relation between them. One can always set what the notation suggests, which is:
\beq
Q_{\mu\al\be} = e^a\mbox{}_\al e^b\mbox{}_\be Q_{\mu ab}\,. \label{Q1Q2}
\eeq
It should be stressed that the above relation is NOT necessary. It can, of course, be put by hand, but no one is
obliged to set this relation, which is an interesting logic issue: observe that the equation (\ref{Q1Q2}) does
not follow from definitions (\ref{defQ1}) and (\ref{defQ2}), but is only suggested by notation. For the sake
of comparison, let us mention that the inverse happens with the standard definition of these notations themselves. We
define, for example, the objects $A_\al$ and $A_a$ such that
\beq
A_\al = e^a\mbox{}_\al A_a\, .  \label{AA}
\eeq
In this case, equation (\ref{AA}) comes unseparated from the definitions of $A_\al$ and $A_a$ (actually, the above
equation is the definition of $A_\al$ or $A_a$ in terms of the definition of the other one). In the case of
non-metricities, the objects $Q_{\mu\al\be}$ and $Q_{\mu ab}$ was not defined by (\ref{Q1Q2}) at all! One can always adopt (\ref{Q1Q2}) for gaining simplicity, but we would like to point out, for the first time, the possibility to deal with two independent kinds of non-metricity.

\subsection{Calculation of $\tilde{\nabla}_\mu e^a\mbox{}_\al$}

As all the objects with a Latin and a Greek label, the covariant derivative of $e^a\mbox{}_\al$
shall necessarily have the corresponding connection for the Latin label, the spin connection,
and the one for Greek label, the affine connection:
\beq
\tilde{\nabla}_\mu e^a\mbox{}_\al =
\partial_\mu e^a\mbox{}_\al - \tilde{\om}^a\mbox{}_{c\mu}\,e^c\mbox{}_\al -
\tilde{\Ga}^\rho\mbox{}_{\al\mu}\,e^a\mbox{}_\rho\,. \label{covderivvierbein}
\eeq
From equation (\ref{defQ1}),
one can write the connection with non-metricity and torsion in terms of the connection
with only torsion, as follows:
\beq
\tilde{\Ga}^\rho\mbox{}_{\al\mu} = \Ga^\rho\mbox{}_{\al\mu} + N^\rho\mbox{}_{\al\mu}\,,
\label{connection}
\eeq
where $N^\rho\mbox{}_{\al\mu} = (Q^\rho\mbox{}_{\al\mu} - Q_\al\mbox{}^\rho\mbox{}_\mu
- Q_\mu\mbox{}^\rho\mbox{}_\al)/2$ and, as usual, $\Ga^\rho\mbox{}_{\al\mu} = \{ ^\rho _{\al\mu}\} +
K^\rho\mbox{}_{\al\mu}$, where $\{ ^\rho _{\al\mu}\}$ is the Christoffel symbol and $K^\rho\mbox{}_{\al\mu}$ is the
contortion tensor\footnote{The torsion tensor is defined by $T^\la\mbox{}_{\al\be} = \tilde{\Ga}^\la\mbox{}_{\al\be}
- \tilde{\Ga}^\la\mbox{}_{\be\al}$.},
$K^\rho\mbox{}_{\al\mu} = (T^\rho\mbox{}_{\al\mu} - T_\al\mbox{}^\rho\mbox{}_\mu
- T_\mu\mbox{}^\rho\mbox{}_\al)/2$. Taking equation (\ref{symspinconnection}) into
account (we mean $\tilde{\om}_{ab\mu} = Q_{\mu ab}/2 + \om_{ab\mu}$), we finally achieve, substituting the above equation into equation
(\ref{covderivvierbein}), and also using $\nabla_\mu e^a\mbox{}_\al = 0$,
\beq
\tilde{\nabla}_\mu e^a\mbox{}_\al = -\frac{1}{2}Q_\mu\mbox{}^a\mbox{}_b\,e^b\mbox{}_\al
-N^\rho\mbox{}_{\al\mu}\,e^a\mbox{}_\rho \,.
\eeq

From the above formula, one can calculate also another derivatives, $\tilde{\nabla}_\mu e^{a\al}$,
$\tilde{\nabla}_\mu e_{a\al}$ and $\tilde{\nabla}_\mu e_a\mbox{}^\al$, which must be done carefully,
keeping in mind that, for example, $\tilde{\nabla}_\mu e^a\mbox{}_\al \neq \eta^{ab}\tilde{\nabla}_\mu e_{b\al}
\neq g_{\al\be}\tilde{\nabla}_\mu e^{a\be}$.

\subsection{Non-trivial vacuum solutions}

One should of course pay attention to $\tilde{\nabla}_\mu \ga^a$, as it is directly related to
quantities like the spinor covariant derivative, $\tilde{\na}_\mu \psi$, required to formulate interaction between matter and geometric variables. Let us consider, for now, the vacuum solutions (only geometric quantities without matter). For this purpose, the most simple action is
\beq
S = -\frac{1}{8\pi G}\int\sqrt{-g}d^4x \tilde{R}\,,
\eeq
where $\tilde{R}$ is the curvature scalar obtained by index contractions of the total curvature tensor
\beq
\tilde{R}^\rho\mbox{}_{\la\mu\nu} = \tilde{\Ga}^\rho\mbox{}_{\la\nu ,\mu} +
\tilde{\Ga}^\rho\mbox{}_{\ta\mu}\tilde{\Ga}^\ta\mbox{}_{\la\nu} - (\mu\leftrightarrow\nu)\,,
\eeq
and the index after the dot means (ordinary) partial derivative introducing the same index. The connections above are the total connections, given by (\ref{connection}). Substituting these connections, we get
\beq
\tilde{R}^\rho\mbox{}_{\la\mu\nu} =  \breve{R}^\rho\mbox{}_{\la\mu\nu} + M^\rho\mbox{}_{\la\nu ||\mu} -
M^\rho\mbox{}_{\la\mu ||\nu} + M^\rho\mbox{}_{\ta\mu}M^\ta\mbox{}_{\la\nu} - M^\rho\mbox{}_{\ta\nu}M^\ta\mbox{}_{\la\mu}\,
\eeq
where $\breve{R}^\rho\mbox{}_{\la\mu\nu}$ is the Riemannian curvature, also $M^\rho\mbox{}_{\mu\nu} = K^\rho\mbox{}_{\mu\nu} + N^\rho\mbox{}_{\mu\nu}$ and the double bar means Riemannian covariant derivative (constructed with Riemannian connection).

The equations of motion for torsion and non-metricity can be achieved (neglecting the surface terms) by variation of action with respect to contortion, $K^\ga\mbox{}_{\al\be}$, and the tensor $N^\ga\mbox{}_{\al\be}$, yielding, respectively,
\beq
T^\be\mbox{}_{\ga\al} + Q_{[\al\ga]}\mbox{}^\be + \de^\be\mbox{}_{[\al} q_{\ga]} - \de^\be\mbox{}_{[\al} Q_{\ga]} - 2\de^\be\mbox{}_{[\al} T_{\ga]} = 0\,, \label{eqK}
\eeq
and
\beq
Q_\ga\mbox{}^{\al\be} + T^{(\al\be)}\mbox{}_\ga + \frac12 \de_\ga\mbox{}^{(\al} q^{\be)}
- \de_\ga\mbox{}^{(\al} Q^{\be)} - \de_\ga\mbox{}^{(\al} T^{\be)}  +
\frac12 g^{\al\be} (2T_\ga - q_\ga) = 0 \,, \label{eqN}
\eeq
where we use the notation for symmetrization and anti-symmetrization such that $a^{(\mu\nu)} = (a^{\mu\nu} + a^{\nu\mu})/2$ and $a^{[\mu\nu]} = (a^{\mu\nu} - a^{\nu\mu})/2$, and the traces $Q_\al$, $q_\al$ and $T_\al$ are defined as
\beq
T_\al = T^\rho\mbox{}_{\al\rho}\, , \;\;\;\;\;
Q_\al = Q^\rho\mbox{}_{\al\rho}  \;\;\; {\rm and} \;\;\;
q_\al = Q_\al\mbox{}_\rho\mbox{}^\rho \,.
\eeq
By taking traces of equations (\ref{eqK}) and (\ref{eqN}), we arrive at the result
\beq
q^\al = 4 Q^\al\,\;\;\; {\rm and} \;\;\; T^\al = \frac32 Q^\al\,,
\eeq
which can be inserted back in (\ref{eqK}) and (\ref{eqN}), yielding, after some algebraic manipulations,
\beq
Q_{\mu\al\be} & = & g_{\al\be}Q_\mu \\
T^\al\mbox{}_{\mu\nu} & = & \frac12\left(\de^\al_\nu Q_\mu - \de^\al_\mu Q_\nu\right)\,.
\eeq

Thus, the vacuum solution with torsion and non-metricity has non-trivial torsion and non-metricity (although not dynamical, obviously), both expressed in terms of the same 4-vector $Q^\al$ (all other degrees of freedom vanishes). Consider, for example, the Einstein-Cartan action together with minimally coupled Dirac fields. In that case, torsion is non-trivial and is algebrically related to fermions. The axial current is the source of torsion (its pseudo-trace). Similar feature happens in our solution: torsion is the source of non-metricity, or vice-versa: non-metricity is the source of torsion.

Of course the situation can change dramatically if we include fermions, which can give rise to other degrees of freedom. Let us investigate then how Dirac fields couple minimally with geometry.

\section{Calculation of $\tilde{\nabla}_\mu \ga^a$}

The Fock-Ivanenko coefficients are the four matrices $\tilde{\Ga}_\mu$, defined by
\beq
\tilde{\na}_\mu\,\psi = \partial_\mu\,\psi + i\tilde{\Ga}_\mu\,\psi\,,
\;\;\; {\rm and} \;\;\;
\tilde{\na}_\mu\,\bar{\psi} = \partial_\mu\,\bar{\psi} - i\bar{\psi}\,\tilde{\Ga}_\mu\,.
\label{GammaReal}
\eeq
The above equations allow us to write the covariant derivative of the matrix $\psi\bar{\psi}$
as $\tilde{\na}_\mu (\psi\bar{\psi}) = \partial_\mu (\psi\bar{\psi}) +
i[\tilde{\Ga}_\mu\,, \psi\bar{\psi}]$, and from this we conclude that one should include
the commutator $i[\tilde{\Ga}_\mu\,, \bf{M}]$ in the expression for the covariant derivative
of the matrix $\bf{M}$. Thus,\footnote{This equation, (\ref{covderivgamma1}), can also be achieved independently
by the requirement that $\bar{\psi}\ga^a\psi$ behaves as a genuine 4-vector in tangent space:
$\tilde{\na}_\mu (\bar{\psi}\ga^a\psi) = \partial_\mu (\bar{\psi}\ga^a\psi) -
\tilde{\om}^a\mbox{}_{b\mu} (\bar{\psi}\ga^b\psi)$.}
\beq
\tilde{\na}_\mu \ga^a = \partial_\mu \ga^a - \tilde{\om}^a\mbox{}_{b\mu}\,\ga^b +
i[\tilde{\Ga}_\mu\,, \ga^a]\,, \label{covderivgamma1}
\eeq
where $\partial_\mu \ga^a$ was written just for completeness (it vanishes).
At the same time, one can suppose that the quantity $\tilde{\na}_\mu \ga^a$ can be
written in terms of combinations of Dirac matrices $\ga_b$, the commutators $\si_{bc} = i[\ga_b\,, \ga_c]/4$ and also the
4x4 identity $\hat{1}$ ($\ga_5$ and $\ga_5\ga_b$ can be
disregarded because of the parity symmetry of $\tilde{\na}_\mu \ga^a$). So, expansion
in this basis yields
\beq
\tilde{\na}_\mu \ga^a = \tilde{A}_\mu\mbox{}^a\,\hat{1} + \tilde{B}_\mu\mbox{}^{ab}\ga_b +
\tilde{C}_\mu\mbox{}^{abc}\si_{bc}\,, \label{covderivgamma2}
\eeq
where $\tilde{A}_\mu\mbox{}^a$, $\tilde{B}_\mu\mbox{}^{ab}$ and $\tilde{C}_\mu\mbox{}^{abc}$
are tensor components, with $\tilde{C}_\mu\mbox{}^{abc} = - \tilde{C}_\mu\mbox{}^{acb}$.
Substituting equation (\ref{covderivgamma2}) into $\tilde{\na}_\mu (\ga_a\ga^a) = 0$,
we get, after some algebra,
$$
2\tilde{A}_\mu\mbox{}^a\ga_a + \left( 2\tilde{B}_\mu\mbox{}^{(ab)} + Q_\mu\mbox{}^{ab} \right) \ga_a\ga_b +
\tilde{C}_\mu\mbox{}^{abc} (\ga_a \si_{bc} + \si_{bc}\ga_a ) = 0\, .
$$
This implies
\beq
\tilde{A}_\mu\mbox{}^a = 0\, ,\;\;\;
\eta_{ab} \tilde{B}_\mu\mbox{}^{ab} = - \eta_{ab} Q_\mu\mbox{}^{ab}\, \;\; {\rm and} \;\;
\tilde{C}_\mu\mbox{}^{[abc]} = 0\,, \label{ABC}
\eeq
where $\tilde{C}_\mu\mbox{}^{[abc]}$ is the normalized totally antisymmetric combination of
$\tilde{C}_\mu\mbox{}^{abc}$, and we used the identity $\ga_a \si_{bc} + \si_{bc}\ga_a = \ep_{abcd}\ga^5\ga^d$.

\section{Relation between spin connection and the Fock-Ivanenko coefficients}

Let us make the simplest choice, restricting by hand the quantities $\tilde{B}_\mu\mbox{}^{ab}$
and $\tilde{C}_\mu\mbox{}^{abc}$, according to
\beq
\tilde{B}_\mu\mbox{}^{ab} = -Q_\mu\mbox{}^{ab}\;\;\; {\rm and} \;\;\;
\tilde{C}_\mu\mbox{}^{abc} = 0\,. \label{choices}
\eeq
Observe that the above equations are more restrictive than equations (\ref{ABC}). Thus, in this particular case, we have
\beq
\tilde{\na}_\mu \ga^a = -Q_\mu\mbox{}^{ab}\ga_b\,.
\label{covderivgamma3}
\eeq

Now we shall substitute the above expression in the equation (\ref{covderivgamma1}), in order to
find a way to write the Fock-Ivanenko coefficients, $\tilde{\Ga}_\mu$, in terms of the spin connection,
$\tilde{\om}^{ab}\mbox{}_\mu$, and the non-metricity, $Q_\mu\mbox{}^{ab}$ (see in Ref. \cite{adler} a very similar approach for Weyl geometry).

By using equation (\ref{covderivgamma3}) in (\ref{covderivgamma1}) we get then
\beq
-Q_\mu\mbox{}^{ab}\ga_b = - \tilde{\om}^a\mbox{}_{b\mu}\,\ga^b + i[\tilde{\Ga}_\mu\,, \ga^a]\,.
\label{eqGamma}
\eeq
In order to solve this equation for $\tilde{\Ga}_\mu$, consider the expansion of the unknown Fock-Ivanenko coefficients $\tilde{\Ga_\mu}$ in the basis
$\left\{ \hat{1},\, \ga_c ,\, \si_{cd} \right\}$:
\beq
\tilde{\Ga}_\mu = \tilde{D}_\mu\,\hat{1} + \tilde{E}_\mu\mbox{}^a\,\ga_a +
\tilde{F}_\mu\mbox{}^{ab}\,\si_{ab}\,,  \label{Gamma}
\eeq
where $\tilde{F}_\mu\mbox{}^{ab} = - \tilde{F}_\mu\mbox{}^{ba}$.
It is very important that these tensor components are real numbers, because, together with
equations (\ref{GammaReal}), it guarantees that $\bar{\psi}\psi$ is a scalar:
$\tilde{\nabla}_\mu(\bar{\psi}\psi) = \partial_\mu(\bar{\psi}\psi)$.
Substituting the above expansion into equation (\ref{eqGamma}), one can write, after straightforward
algebra,
\beq
\left(-Q_\mu\mbox{}^{ab} + \tilde{\om}^{ab}\mbox{}_\mu - 2\tilde{F}_\mu\mbox{}^{ab}\right) \ga_b
- 4\tilde{E}_\mu\mbox{}^b \eta^{ac}\,\si_{bc} = 0\,, \label{eq}
\eeq
where we have used the identity $[\si_{bc},\ga_a] = i\ga_b\eta_{ac} - i\ga_c\eta_{ab}$.
With respect to Latin indices, remember that $\tilde{F}_\mu\mbox{}^{ab}$ is antisymmetric and
$Q_\mu\mbox{}^{ab}$ is symmetric, so we conclude
\beq
\tilde{F}_\mu\mbox{}^{ab}  =  \frac{1}{2}\tilde{\om}^{[ab]}\mbox{}_\mu\;\;\; {\rm and} \;\;\;
\tilde{E}_\mu\mbox{}^a = 0\,,
\eeq
Notice that equation (\ref{symspinconnection}) is also proven (independently) from (\ref{eq}). It is worth mentioning that $\tilde{D}_\mu$ does not need to be zero. Then (\ref{Gamma}) reads
\beq
\tilde{\Ga}_\mu = \frac{1}{2}\tilde{\om}^{[ab]}\mbox{}_\mu\,\si_{ab} + \tilde{D}_\mu\,\hat{1}\,. \label{FI}
\eeq
It is very interesting that the literature on this subject\footnote{See, for instance, the works in Refs. \cite{adler,fariborz,adak,formiga,latorre}.} point to the presence of the arbitrary 4-vector $\tilde{D}_\mu$.
The first term, proportional to $\si_{ab}$, is the usual term that appears in the covariant derivative of a spinor under the minimal coupling prescription. Since all 4-vectors from the geometric content refer to traces of torsion and non-metricity, it is natural to consider these traces (or its combinations) as good candidates for the quantity $\tilde{D}_\mu$.

\section{The arbitrariness of the Fock-Ivanenko coefficients and $\tilde{\nabla}_\mu \ga^a$}

If we did not make the very restrictive choices (\ref{choices}), equation (\ref{covderivgamma3}) would have to be rewritten as
\beq
\tilde{\na}_\mu \ga^a = \tilde{B}_\mu\mbox{}^{ab}\ga_b + \tilde{C}_\mu\mbox{}^{abc}\si_{bc}\,,
\label{covderivgamma4}
\eeq
where $\tilde{B}_\mu\mbox{}^{ab}$ and $\tilde{C}_\mu\mbox{}^{abc}$ satisfy the necessary conditions (\ref{ABC}). This means that the covariant derivative of $\ga^a$ is actually much more arbitrary throughout a second rank tensor ($\tilde{B}_\mu\mbox{}^{ab}$) with arbitraries anti-symmetric and traceless parts and a third rank tensor satisfying $\tilde{C}_\mu\mbox{}^{[abc]} = 0$.

Now, substituting (\ref{covderivgamma4}) into (\ref{covderivgamma1}) together with the expansion (\ref{Gamma}), one is able to find the following result:
\beq
\tilde{\Ga}_\mu = \tilde{D}_\mu\,\hat{1} + \frac{1}{6}\tilde{C}^{ba}\mbox{}_b\,\ga_a +
\frac{1}{2}\left( \tilde{B}_\mu\mbox{}^{[ab]} + \tilde{\om}^{[ab]}\mbox{}_\mu \right)\,\si_{ab}\,. \label{Ga}
\eeq
In this equation, the arbitrariness of $\tilde{\Ga}_\mu$ (except $\tilde{D}_\mu$) comes from the arbitrariness of
$\tilde{\nabla}_\mu \ga^a$. This arbitrariness can not be eliminated or reduced by some additional condition besides
$\tilde{\na}_\mu (\bar{\psi}\ga^a\psi) = \partial_\mu (\bar{\psi}\ga^a\psi) - \tilde{\om}^a\mbox{}_{b\mu} (\bar{\psi}\ga^b\psi)$. For example, if one considers
\beq
\tilde{\na}_\mu (\bar{\psi}\ga^a\ga^b\psi) = \partial_\mu (\bar{\psi}\ga^a\ga^b\psi) -
\tilde{\om}^a\mbox{}_{c\mu} (\bar{\psi}\ga^c\ga^b\psi) - \tilde{\om}^b\mbox{}_{c\mu} (\bar{\psi}\ga^a\ga^c\psi)\,,
\eeq
no more and no less than equations (\ref{ABC}) would be derived likewise.

\section{Conclusions: the case with metricity}

The notations we have used are very convenient for investigating the case with torsion alone. All quantities with tilde are defined in the spacetime with non-metricity. For the case with metricity and non-zero torsion, the quantities is written without tilde, such that all equations will be essentially the same, with the obvious feature $Q_\mu\mbox{}^{ab} = 0$. Equation
(\ref{Ga}), for example, would read
\beq
\Ga_\mu = D_\mu\,\hat{1} + \frac{1}{6}C^{ba}\mbox{}_b\,\ga_a +
\frac{1}{2}\left( B_\mu\mbox{}^{[ab]} + \om^{ab}\mbox{}_\mu \right)\,\si_{ab}\,.
\eeq
Here, there are also those arbitrariness found in the non-metricity case. It is interesting to observe that the alleged reason by which there should be a 4-vector $\tilde{D}_\mu$ in equation (\ref{FI}) works perfectly well in saying that this 4-vector is also present in the case with just torsion. The other quantities, $C^{ba}\mbox{}_b$ and $B_\mu\mbox{}^{[ab]}$, can also be present in the expression of $\Ga_\mu$ without contradiction, just as it happens in the non-metricity case, $\tilde{\Ga}_\mu$. I haven't seen any consistency condition that can rule out those arbitrariness.

\section*{Acknowledgments}
G.B.P. is grateful to CNPq and FAPEMIG for partial support. The author acknowledges Prof. I. Shapiro (UFJF) and Mr. A. F. Andrade (IFMG) for fruitful discussions in the beginning of this work.



\begin{thebibliography}{99}

\bibitem{hehl} F. W. Hehl, P. Von Der Heyde, G. D. Kerlick and J. M. Nester, {\it Rev. Mod. Phys.} {\bf 48}: 393-416, 1976.
DOI: 10.1103/RevModPhys.48.393

\bibitem{adler} R. J. Adler, {\it J. Math. Phys.} {\bf 11}: 1185, 1970. DOI: 10.1063/1.1665246 	

\bibitem{poberii} E. A. Poberii, {\it Gen. Rel. Grav.} {\bf 26} n. 10: 1011-1054, 1994.

\bibitem{shapiro} I.L. Shapiro, {\it Phys. Rept.} {\bf 357}: 113-213, 2002.

\bibitem{saridakis} Yi-Fu Cai, Salvatore Capozziello, Mariafelicia De Laurentis and Emmanuel N. Saridakis, {\it Rept. Prog. Phys.} {\bf 79}  no.10: 106901, 2016. ArXiv: 1511.07586 [gr-qc].

\bibitem{salim1} J. M. Salim and S. L. Saut\'u, {\it Class. Quantum Grav.} {\bf 13}: 353-360, 1996.
DOI: 10.1088/0264-9381/13/3/004

\bibitem{salim2} H. P. de Oliveira, J. M. Salim and S. L. Saut\'u, {\it Class. Quantum Grav.} {\bf 14}: 2833-2843, 1997.
DOI: 10.1088/0264-9381/14/10/010

\bibitem{aguila} R. Aguila, J. E. M. Aguilar, C. Moreno and M. Bellini, {\it Eur. Phys. J.} {\bf C74} n.11: 3158, 2014.
DOI: 10.1140/epjc/s10052-014-3158-y

\bibitem{fariborz} A. H. Fariborz and D. G. C. McKeon, {\it Class. Quant. Grav.} {\bf 14}: 2517-2525, 1997.
DOI: 10.1088/0264-9381/14/9/009.

\bibitem{baburova} O. V. Baburova, B. N. Frolov and R. S. Kostkin, e-Print: arXiv:1006.4761 [gr-qc]

\bibitem{koivisto} J. B. Jimenez and T. S. Koivisto, {\it Phys. Lett.} {\bf B756}: 400-404, 2016.
DOI: 10.1016/j.physletb.2016.03.047

\bibitem{moffat} Marco de Cesare, John W. Moffat and Mairi Sakellariadou, {\it Eur.Phys.J.} {\bf C77} no.9: 605, 2017.
DOI: 10.1140/epjc/s10052-017-5183-0

\bibitem{adak} M. Adak, T. Dereli and L.H. Ryder {\it Int.J.Mod.Phys.} {\bf D12}: 145-156, 2003.
DOI: 10.1142/S0218271803002445.

\bibitem{formiga} J.B. Formiga and C. Romero, {\it Int. J. Geom. Meth. Mod. Phys.} {\bf 10}: 1320012, 2013.
DOI: 10.1142/S0219887813200120.

\bibitem{olmo} G. J. Olmo and D. Rubiera-Garcia, {\it J. Phys. Conf. Ser.} {\bf 600} n.1: 012041, 2015.
DOI: 10.1088/1742-6596/600/1/012041
	
\bibitem{koste} J. Foster, V. Alan Kosteleck\'y and R. Xu, {\it Phys. Rev.} {\bf D95} no.8: 084033, 2017.
DOI: 10.1103/PhysRevD.95.084033

\bibitem{snow} R. Lehnert, W.M. Snow, Z. Xiao and R. Xu, {\it Phys. Lett.} {\bf B772}: 865-869, 2017.
DOI: 10.1016/j.physletb.2017.07.059

\bibitem{latorre} A. Delhom I Latorre, Gonzalo J. Olmo and Michele Ronco, Phys. Let. {\bf B780}: 294-299. E-Print arXiv:1709.04249. DOI: 10.1016/j.physletb.2018.03.002

\bibitem{stg} Muzaffer Adak, Ozcan Sert, Turk.J.Phys. {\bf 29}: 1-7, 2005; Muzaffer Adak, M. Kalay, Ozcan Sert, Int.J.Mod.Phys. {\bf D15}: 619-634, 2006 DOI: 10.1142/S0218271806008474; Muzaffer Adak, Ozcan Sert, Mestan Kalay, Murat Sari, Int.J.Mod.Phys. {\bf A28}: 1350167, 2013; J. B. Jim\'enez, L. Heisenberg and T. S. Koivisto, {\it Teleparallel Palatini Theories}, preprint arXiv: 1803.10185.

\bibitem{review} Lavinia Heisenberg, {\it A systematic approach to generalizations of General Relativity and their cosmological implications}, preprint arXiv: 1807.01725.

\bibitem{pono} V.N. Ponomariov and Yu. Obukhov {\it Gen. Rel. Grav.} {\bf 14}: 309, 1982.

\bibitem{neeman} Yuval Ne'eman, {\it Ann. Inst. Henri Poincar\'e}, Section A {\bf 28} n.4: 369-378, 1978.

\bibitem{hurley} D.J. Hurley and M.A. Vandyck, {\it J. Phys. A: Math. Gen.} {\bf 27}: 4569-4580, 1994.
DOI: 10.1088/0305-4470/27/13/030

\bibitem{fatibene} Lorenzo Fatibene, Marco Ferraris, Mauro Francaviglia and Marco Godina,
Presented at 6th International Conference on Differential Geometry and Applications (C95-08-28.9). e-Print: gr-qc/9608003.

\end{thebibliography}
\end{document}